\documentclass[conference]{IEEEtran}

\usepackage{multirow}
\ifCLASSINFOpdf
   \usepackage[pdftex]{graphicx}
   
   \graphicspath{{./pdf/}{./jpeg/}}
   \DeclareGraphicsExtensions{.pdf,.jpeg,.png}
\else
   \usepackage[dvips]{graphicx}
   \graphicspath{{./eps/}}
   \DeclareGraphicsExtensions{.eps}
\fi
\usepackage{mathtools}
\usepackage{color}
\usepackage{algorithm}
\usepackage{algpseudocode}

\usepackage{times}
\usepackage{epsfig}
\usepackage{graphicx}
\usepackage{amsmath}
\usepackage[utf8]{inputenc}
\usepackage{amssymb}
\usepackage{romannum}
\usepackage{color}
\usepackage{caption}
\usepackage{subcaption}
\usepackage{float}

\usepackage{float}

\usepackage[margin=1in]{geometry}  
\usepackage{array}                
\usepackage{booktabs} 
\usepackage[table,xcdraw]{xcolor}

\usepackage{tabularx}

\usepackage{algorithm,algpseudocode}

\usepackage{url}

\usepackage[colorlinks=true, linkcolor=blue, citecolor=blue, urlcolor=blue]{hyperref}

\begin{document}
%
\title{Knowledge Acquisition on Mass-shooting Events via LLMs for AI-Driven Justice}




\author{\IEEEauthorblockN{Benign John Ihugba$^{3}$, Afsana Nasrin$^{1}$, Ling Wu$^{2}$, Lin Li$^{3}$, Lijun Qian$^{1}$, and Xishuang Dong$^{1}$}
\IEEEauthorblockA{$^{1}$Department of Electrical and Computer Engineering, $^{2}$College of Juvenile Justice, $^{3}$Department of Computer Science\\ 
Prairie View A\&M University, Prairie View, TX 77446, USA \\
Email: bihugba@pvamu.edu, anasrin@pvamu.edu, liwu@pvamu.edu,\\ lilin@pvamu.edu, liqian@pvamu.edu,  xidong@pvamu.edu}
}


\maketitle

\begin{abstract}
Mass-shooting events pose a significant challenge to public safety, generating large volumes of unstructured textual data that hinder effective investigations and the formulation of public policy. Despite the urgency, few prior studies have effectively automated the extraction of key information from these events to support legal and investigative efforts. This paper presented the first dataset designed for knowledge acquisition on mass-shooting events through the application of named entity recognition (NER) techniques. It focuses on identifying key entities such as offenders, victims, locations, and criminal instruments, that are vital for legal and investigative purposes. The NER process is powered by Large Language Models (LLMs) using few-shot prompting, facilitating the efficient extraction and organization of critical information from diverse sources, including news articles, police reports, and social media. Experimental results on real-world mass-shooting corpora demonstrate that GPT-4o is the most effective model for mass-shooting NER, achieving the highest Micro Precision, Micro Recall, and Micro F1-scores. Meanwhile, o1-mini delivers competitive performance, making it a resource-efficient alternative for less complex NER tasks. It is also observed that increasing the shot count enhances the performance of all models, but the gains are more substantial for GPT-4o and o1-mini, highlighting their superior adaptability to few-shot learning scenarios.

\end{abstract}

\begin{IEEEkeywords} Knowledge Acquisition, Mass-shooting Events, Large Language Models (LLMs), AI-Driven Justice; \end {IEEEkeywords}

%
\IEEEpeerreviewmaketitle

\section{Introduction}
\label{sec1}

AI-driven justice refers to the use of AI tools  and  models to improve fairness, efficiency, and consistency in legal processes~\cite{de2024advancing}. These range from automated evidence analysis to real-time crime reporting systems, thus supporting near-real-time law enforcement, judicial decision-making, and policymaking. For instance, the Indianapolis Metropolitan Police Department collaborated with a team of AI researchers to revisit a 20-year-old unsolved murder case. According to the Indianapolis Star, the team used an AI tool to analyze $3,000$ pages of investigative files, including handwritten notes, to create a timeline of events. Similarly, a Miami Police Department official testified during a U.S. Senate subcommittee hearing that facial recognition technology and AI tools have significantly contributed to reducing homicide and violent crime rates in Miami in recent years~\cite{gain2024traction}. AI-powered approaches have demonstrated the ability to enhance the capture and synthesis of vast amounts of information by quickly identifying relationships, patterns, and anomalies that might be overlooked by human analysts.

\begin{figure*}[!h]
    \centering
    \includegraphics[width=1\linewidth]{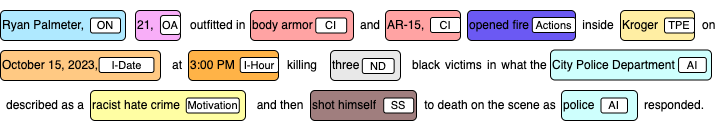}
    \caption{An example of NER on a mass-shooting event for knowledge acquisition. It includes various types of named entities including Offender Name (ON), Offender Age (OA), Criminal Instrument (CI), Actions, Type of Premises Entered (TPE), Incident Date (I-Date), Incident Hour (I-Hour), Number of Deaths (ND), Agency Information (AI), Motivation, Shooter Suicidality (SS).}
    \label{fig_ner}
\end{figure*}

A Knowledge Graph (KG) is a structured representation of knowledge including various entities—such as suspects, victims, and locations—and the relationships between the entities~\cite{9416312}, which captured complex, interconnected information from complex and massive data by various AI techniques, including inference engines and predictive models~\cite{graphAware2023data}. In this process, knowledge acquisition is the key to the success, that aims to extract, structure, and validate information from multiple data sources, where named entity recognition (NER) is an important foundation of knowledge acquisition that recognizes and classifies named entities (e.g., offender, victim, organization, location) into predefined categories~\cite{wang2023gpt}.

This study focused on knowledge acquisition on mass-shooting events via NER techniques to prompt AI-driven justice. The mass-shooting events represent a significant public safety crisis worldwide. For instance, in 2025, at least 14 people lost their lives, and dozens were injured when a driver drove a pickup truck into a crowd during New Year’s celebrations on Bourbon Street in New Orleans\footnote{https://www.cnn.com/us/live-news/new-orleans-mass-casualty-bourbon-street-01-01-25-hnk/index.html}. Such events often unfold rapidly and can yield extensive media coverage, law enforcement investigations, and public discourse. When analyzing such events which have complex social, psychological, and legal dimensions, the data generated (e.g., eyewitness reports, official statements and media articles) is usually voluminous, unstructured and found in multiple sources. A critical challenge lies in transforming these fragmented data points into a coherent representation that can be used to support investigations, policy-making, and predictive analyses \cite{zhao-etal-2020-knowledge, devlin2018bert}. For instance, between 2006 and 2016, the United States recorded an annual average of over $33,000$ gun-related deaths, prompting calls to treat gun violence as a public health issue and urging news media to adopt more holistic reporting practices~\cite{silverman2018publicHealth}.

In this paper, we leverage LLM-based NER using few-shot learning to extract entities from real-world mass shooting events. A mass shooting dataset comprising 153 events was first constructed, with data sourced from MotherJones\footnote{https://www.motherjones.com/politics/2012/12/mass-shootings-mother-jones-full-data/}. The dataset is categorized into four major entity types—Offender, Victim, Environment, and Justice Agency—further divided into 41 subcategories to capture detailed information about mass shooting events. We employed several LLMs, including GPT-3.5, GPT-4o, and GPT-o1-mini, to identify these entities. Experimental results on real-world mass-shooting corpora show that GPT-4o is the most effective model for mass-shooting NER, achieving the highest Micro Precision, Micro Recall, and Micro F1-scores. In contrast, o1-mini demonstrates competitive performance, serving as a resource-efficient alternative for less complex NER tasks. The results further indicate that increasing the shot count enhances the performance of all models, with the improvement being more substantial for GPT-4o and o1-mini, underscoring their superior adaptability to few-shot learning scenarios.

Our contributions are summarized as:

\begin{itemize}
\item As far as we know, it is the first dataset for knowledge acquisition on mass shooting events using named entity recognition (NER) techniques. This dataset not only fosters AI-driven justice by providing structured insights into such incidents but also enhances the applicability of LLMs for NER tasks, paving the way for more robust and context-aware information extraction.

\item We conducted a systematic validation and examination of GPT-based LLMs for NER on mass shooting events. Experiment results provide valuable insights into the performance, limitations, and adaptability of LLMs for processing data in low-resource domains. These findings underscore the potential of LLMs to improve NER tasks where data availability is limited, offering a scalable approach to extract and analyze critical information in specialized contexts like mass shootings.

\end{itemize}

\section{Task Definition }
\label{sec2}
\begin{table*}[h]
    \centering
    \caption{Types of named entities and their description related to offense, offender, and victim involved on mass-shooting events.}
    \begin{tabular}{|p{5cm}|p{11cm}|}
    \hline
     \multicolumn{2}{|c|}{\textbf{Offense Information}} \\
           \hline
	 \textbf{Type of Named Entity} & \textbf{Description} \\
        \hline
        \textbf{I-Date (Incident Date)} & Date of the incident \\
        \hline
        \textbf{I-Day (Incident Day)} & Weekday (Monday) Weekends (Saturday) \\
        \hline
        \textbf{I-Hour (Incident Hour)} & Hour that the incident happened \\
        \hline
        \textbf{GPE (geo-political entity)} & Countries, cities, states. \\
        \hline
        \textbf{Address} & Address of the incident \\
        \hline
        \textbf{TPE (Type of Premises Entered)} & Workplace, government, retail, bar/restaurant, residential location, outdoors, school, university, church, etc. \\
        \hline
        \textbf{Actions} & Actions or movements that occurred in the shooting event such as gunfire, fleeing, hiding, evacuation, assisting others, falling, communication, confusion, law enforcement response, etc. \\
        \hline
        \textbf{CI (Criminal Instrument)}  & Instruments used: handguns, assault rifles, bulletproof vests, knives, etc. \\ 
         \hline
	\textbf{AF (Access to Firearms)} & How the perpetrator accessed the weapon: legal/illegal purchase, stole guns, received from family, etc. \\
	\hline
	\textbf{OSS (Offender suspected use)} & Offender suspected use of alcohol or drugs \\
	\hline
	\textbf{MHS (Mental Health issue)} & Any mental health issues involved \\
	\hline
	\textbf{SS (Shooter Suicidality)} & Any shooter suicide \\
	\hline
	\textbf{Motivation} & Revenge related to employment, or seeking fame/attention/notoriety \\
	\hline
	\textbf{V (Verdict)} & Formal court decision on guilt or innocence of the defendant \\ \hline
	\hline

	\multicolumn{2}{|c|}{\textbf{Offender Information}} \\
	\hline
	 \textbf{Type of Named Entity} & \textbf{Description} \\
	\hline
	\textbf{ON (Offender Name)}  & Name of the offender \\
	\hline
	\textbf{OA (Offender Age)} & Age of the offender \\ \hline
	\textbf{OS (Offender Sex)} & Sex of the offender \\ \hline
	\textbf{OR (Offender Race)} 	& Race of the offender \\ \hline
	\textbf{OE (Offender Ethnicity)} & Ethnicity of the offender \\ \hline
	\textbf{OO (Offender Occupation)}  & Police, teacher, professor, singer, dancer, etc. \\ \hline
	\textbf{WS (Warning Signs)} & Online/offline behaviors suggesting a potential mass shooting \\ \hline
	\textbf{OPCR (Offender prior criminal record)} & Any prior criminal record \\ \hline
	\textbf{OHV (Offender history of violence)} & Domestic violence or any other form of violence \\ \hline
	\textbf{OMB (Offender military background)} & Any information on military background \\ \hline
	\textbf{OOR (Offender Organization)}  & Any organizations affiliated with the offender (not primary occupation) \\ \hline
	\hline

	\multicolumn{2}{|c|}{\textbf{Victim Information}} \\
	\hline
	 \textbf{Type of Named Entity} & \textbf{Description} \\
	 \hline
	\textbf{VC (Victim count)} & Number of victims \\ \hline
	\textbf{VN (Victim Name)} & Name of the victim \\ \hline
	\textbf{VA (Victim Age)} & Age of the victim \\ \hline
	\textbf{VS (Victim Sex)} & Sex of the victim \\ \hline
	\textbf{VR (Victim Race)} & Race of the victim \\ \hline
	\textbf{IT (Injury Types)} & Injuries such as gunshot wounds, cuts, etc. \\ \hline
	\textbf{NI (Number of injuries)} & Number of victims injured (but not killed) \\ \hline
	\textbf{ND (Number of deaths)} & Number of victims killed \\ \hline
	\textbf{RO (Relationships to Offenders)} & Relationship(s) between victims and offenders \\ \hline
	\textbf{VOC (Victim Occupation)} & Police, teacher, professor, singer, dancer, etc. \\ \hline
	\textbf{VOR (Victim Organization)} & Companies, agencies, institutions, universities, etc. \\ 
	\hline
	
    \end{tabular}

    \label{tab:offense}
\end{table*}

This study focuses on knowledge acquisition for AI-driven justice by applying named entity recognition (NER) to mass shooting events. A named entity is a word or phrase that distinctly identifies an item within a set of similar attributes~\cite{sharnagat2014named}, such as organization or location names in general domains. In this study, NER involves detecting and classifying entities in mass shooting-related text into predefined categories. Given a token sequence $s = <w_1, w_2, .., w_N>$, NER outputs tuples $<I_s, I_e, t>$, where $I_s \in [1, N]$ and $I_e \in [1, N]$ denote the entity’s start and end tokens mentioned in the sequence $s$, and $t$ represents its category. Figure~\ref{fig_ner} illustrates an example where an NER system identifies various named entities in a summary of a specific mass shooting event.

\begin{table*}[h]
    \centering
    \caption{Types of named entities and their description related to environment and agency involved in  mass-shooting events.}
    \begin{tabular}{|p{5cm}|p{11cm}|}
    \hline
	\multicolumn{2}{|c|}{\textbf{Environmental or Contextual Factors}} \\
	  \hline
	 \textbf{Type of Named Entity} & \textbf{Description} \\
	\hline
	\textbf{Location} & Non-GPE locations (mountains, bodies of water, etc.) \\ \hline
	\textbf{NOE (Non-Offense Event)} & Named hurricanes, battles, wars, sports events, etc. \\ \hline
	\textbf{Quantity} & Numerical values or quantities (1, 3.5, two, one and a half, etc.) \\ \hline
	\textbf{ET (Event Terminator)}  & Factors that terminate the event (e.g., police intervention) \\ \hline
	\hline

	\multicolumn{2}{|c|}{\textbf{Justice Agency}} \\
	\hline
	\textbf{Type of Named Entity} & \textbf{Description} \\ \hline
	\textbf{AI (Agency Information)} 	& Justice agencies/employees who responded (police, SWAT, fire department, etc.) \\ 
	\hline

    \end{tabular}

    \label{tab:envi}
\end{table*}

\section{Data Collection and Annotation}

\subsection{Data Collection and Preparation}
The data for this study is collected from \textbf{Mother Jones mass-shooting dataset}\footnote{https://www.motherjones.com/politics/2012/12/mass-shootings-mother-jones-full-data/}, which offers comprehensive documentation of mass shootings in the United States. We extracted event summaries for NER and utilized the Natural Language Toolkit (NLTK)\footnote{https://www.nltk.org} to tokenize these summaries, facilitating knowledge acquisition through NER.

\subsection{Data Annotation}

In collaboration with justice domain experts, we establish detailed annotation guidelines following the standard BIO format. Additionally, under expert supervision, two students undergo training to annotate the data through three iterative rounds below.

\begin{itemize}
\item \textbf{Initial Labeling:} Annotators independently annotated documents adhering to the standard BIO formatting using a set of predefined entity types.

\item \textbf{Cross-Verification:} Random subsets of annotated documents were cross-checked by annotators to identify inconsistencies or ambiguous interpretations.

\item \textbf{Adjudication Rounds:} Discrepancies were resolved through consensus-based discussions, and the annotation guidelines were refined accordingly.

\end{itemize}

\begin{figure*}
    \centering
    \includegraphics[width=1\linewidth]{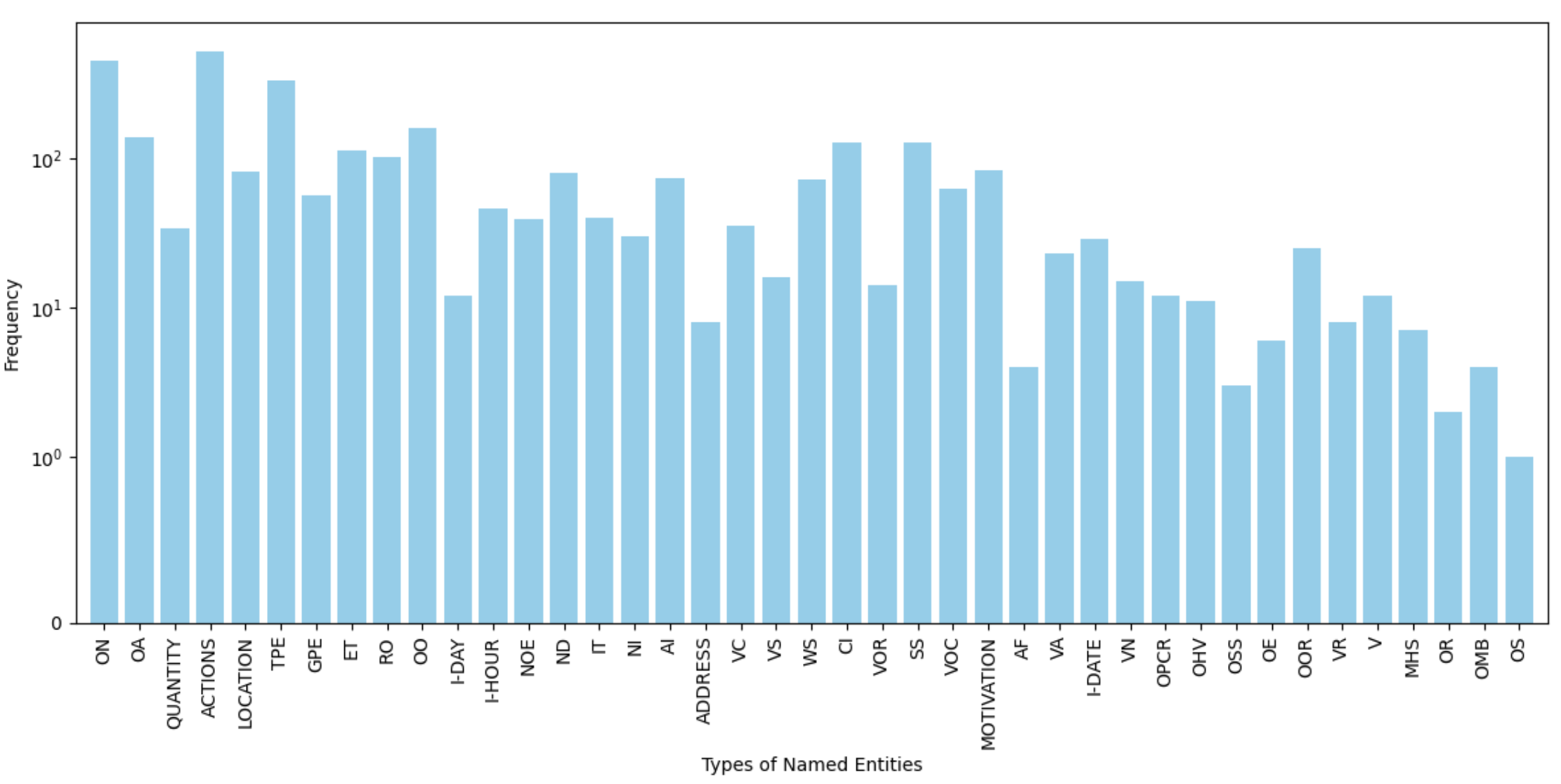}
    \caption{Distributions of named entities involved in mass-shooting events.}
    \label{fig:dist}
\end{figure*}

Tables~\ref{tab:offense} and \ref{tab:envi} provide detailed information on the types of named entities involved in mass-shooting events, representing the extracted knowledge for acquisition. Moreover, Figure~\ref{fig:dist} presents the distributions of named entities involved in mass-shooting events. It is observed that named entity distribution is skewed into favouring frequently reported categories such as Offender Name, Actions, Age and Type of premises entered (TPE) while more other entities such as Offender military background, Mental Health issues, Verdict, which appears less frequently provided crucial specificity. 

\section{LLMs and Prompt Engineering}
\label{sec3}

LLMs, such as the GPT series, have demonstrated remarkable performance in NER~\cite{wang2023gpt}. Pretrained on vast textual corpora, LLMs capture rich linguistic and contextual patterns, enabling zero- to few-shot adaptation to specialized domains, such as domain-specific knowledge acquisition events in this study. The GPT series has emerged as a leading choice due to its strong reasoning capabilities, contextual awareness, and adaptability, outperforming many state-of-the-art LLMs. Furthermore, few-shot prompting enables LLMs to utilize annotation guidelines through only a minimal number of labeled examples~\cite{bhandari2023survey}. This few-shot approach significantly reduces the annotation workload, accelerates model deployment, and enhances the flexibility and scalability of the NER pipeline. In this study, we leverage GPTs with few-shot prompting to perform NER on mass-shooting event summaries.

\subsection{LLMs}

This study leverages three OpenAI LLMs, including GPT-3.5, GPT 4o, and o1-mini, for NER on mass-shooting event summaries

\begin{itemize}
\item GPT-3.5 is a widely recognized generation of LLMs that exhibits notable improvements in contextual reasoning and coherent responses. It excels at understanding instructions and also maintains conversation-like flows, which makes it suitable for tasks where subtle narrative elements need to be identified~\cite{liu2023evaluating}.
However, it tends to produce plausible but incorrect or nonsensical answers and hallucinations~\cite{chatgpt35overview}. It is also known to hallucinate, generating information not present in training data~\cite{aimlapi2025comparison}. It lacks advanced tools such as web browsing, memory functionality, and file uploads~\cite{turtleverse_o1mini_2025}. 

\item GPT-4o is the new iteration in the GPT model family, succeeding GPT-4. It offers enhanced reasoning depth, improved factual consistency, and a stronger ability to infer relationships between entities compared to GPT-3.5. These advancements are potentially beneficial for extracting complex entity types, such as an offender’s history of violence or mental health issues~\cite{enwiki:1273223835}. Additionally, GPT-4o demonstrates greater granularity, fewer hallucinations, and improved handling of long and complex input texts, making it well-suited for tasks requiring precise and context-aware entity recognition~\cite{dua2022successive}.

\item The o1-mini is a smaller, optimized version of the o1 model series, specifically designed to enhance reasoning capabilities in tasks such as math, coding, and cybersecurity. It operates with reduced resource requirements while maintaining strong performance. Previous studies indicate that while o1-mini may not fully match the depth of understanding of the larger o1 model in highly intricate general contexts, it still delivers solid performance for many common entity types~\cite{zhuang2023survey, sun2019mobilebert}. This study aims to evaluate whether o1-mini can achieve near-comparable accuracy and performance in identifying complex, domain-specific entities while consuming fewer computational resources. Such efficiency could potentially enable on-device or edge deployment scenarios, making it a practical choice for resource-constrained environments.

\end{itemize}

\subsection{Prompt Engineering}

\begin{figure}
    \centering
    \includegraphics[width=1\linewidth]{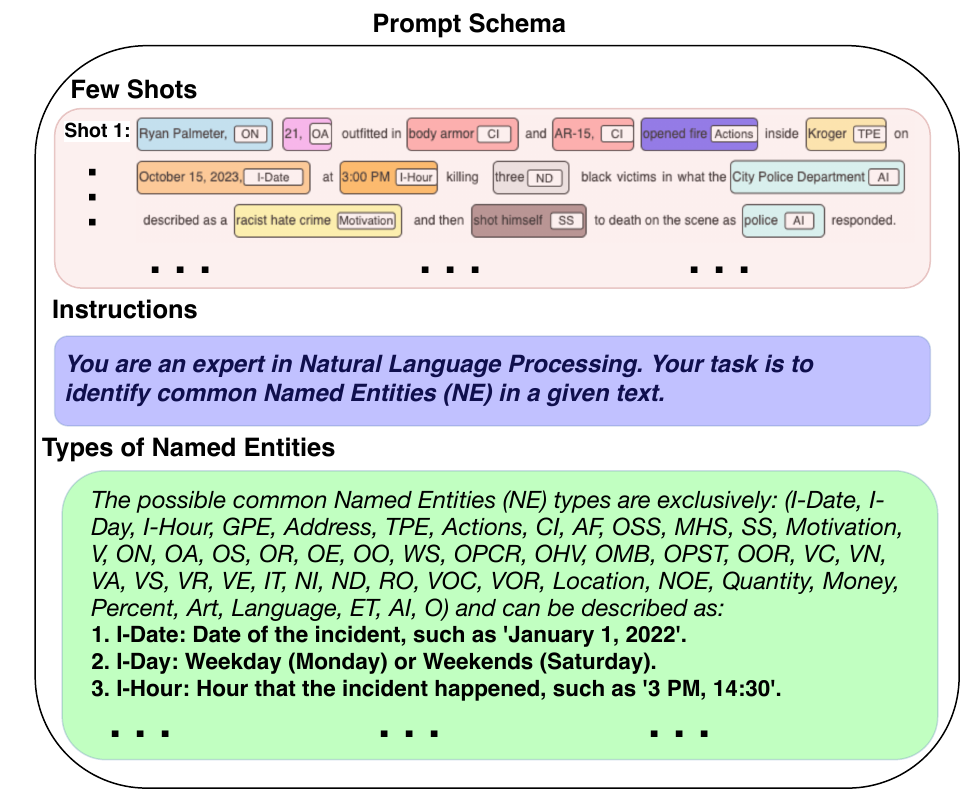}
    \caption{Prompt schema of NER on mass-shooting events.}
    \label{fig:prompt}
\end{figure}

Prompt engineering is vital for shaping LLM behavior to perform few-shot NER on mass-shooting texts. We construct a carefully formatted prompt that includes a handful of in-context exemplars to demonstrate how unstructured reports are converted into structured entity outputs~\cite{gao2020making}.  Figure~\ref{fig:prompt} presents the schema of the prompt in this study.

\begin{itemize}
\item \textbf{Few shots}: Each shot in the prompt illustrates an input snippet paired with the desired output listing key entities, thereby guiding the model to mimic this format.

\item \textbf{Instructions}: It specifies the task completed.

\item \textbf{Types of named entities}: We design the prompt to highlight the target entity types defined in our annotation schema – for instance, Offender Name, Offender Age, Weapon (Criminal Instrument), Type of Premises Entered, etc. using short labels (“ON”, “OA”, “CI”, “TPE”, etc.) as field prefixes. These formatting choices draw inspiration from recent prompt-based NER methods like GPT-NER~\cite{wang2023gpt}, which convert sequence labeling into a generative task with special token cues. 
\end{itemize}

This approach leverages the LLM’s in-context learning ability to internalize annotation guidelines from just a few shots of examples, rather than requiring extensive fine-tuning~\cite{gao2020making}. Such prompt engineering has been shown to be crucial, as LLMs are highly sensitive to prompt design and require clear instructions to perform optimally~\cite{ye2023prompt}.

This prompt-based strategy enhances interpretability and accuracy in several ways. First, by constraining the output to a structured template, we ensure the model’s responses are directly usable for our knowledge graph, each entity is clearly tagged, improving downstream interpretability and consistency. Second, providing schema-focused exemplars helps the LLM attain higher precision and recall on entity extraction, as it reduces off-target or hallucinated outputs~\cite{wang2023gpt}.

\section{Experiments}
\label{sec4}

\subsection{Evaluation Metrics}
Model performance was evaluated using standard NER metrics, including F1-score, Micro Precision, Micro Recall, and Micro F1-score~\cite{li2020survey}. The F1-score provides a balance between precision and recall, where precision measures the accuracy of extracted entities, and recall assesses the coverage of relevant entities. Additionally, Micro Precision, Micro Recall, and Micro F1-score offer a comprehensive summary of the model’s overall performance.

\subsection{Experimental Results and Discussions}

\begin{table}[h]
\centering
\caption{Performance comparison of NER on mass-shooting events with different shots of samples between GPT 3.5, GPT 4o and o1-mini.}

\begin{tabular}{cccc}
\toprule
\multicolumn{4}{c}{Prompt with 5 shots}\\
\hline
\textbf{Model} & \textbf{Micro Precision} & \textbf{Micro Recall} & \textbf{Micro F1-Score} \\
\hline
GPT 3.5   & 0.5950 & 0.3062 & 0.4043 \\
GPT 4o    & \textbf{0.6691} & \textbf{0.6471} & \textbf{0.6579} \\
 o1-mini   & 0.6387 & 0.5741 & 0.6047 \\
 \hline
 \hline
 \multicolumn{4}{c}{Prompt with 10 shots}\\
\hline
GPT 3.5   & 0.6227 & 0.3445 & 0.4436 \\
GPT 4o    & \textbf{0.6903} & \textbf{0.6515} & \textbf{0.6703} \\
 o1-mini   & 0.6458 & 0.5861 & 0.6145 \\
 \hline
 \hline
 \multicolumn{4}{c}{Prompt with 14 shots}\\

\hline
 GPT 3.5   & 0.6150 & 0.3241 & 0.4245 \\
 GPT 4o    & \textbf{0.6831} & \textbf{0.6814} & \textbf{0.6823} \\
 o1-mini   & 0.6556 & 0.6051 & 0.6293 \\
\bottomrule
\end{tabular}

\label{tab:results}
\end{table}

\begin{figure*}[h!]
    \centering
    \includegraphics[width=1\linewidth]{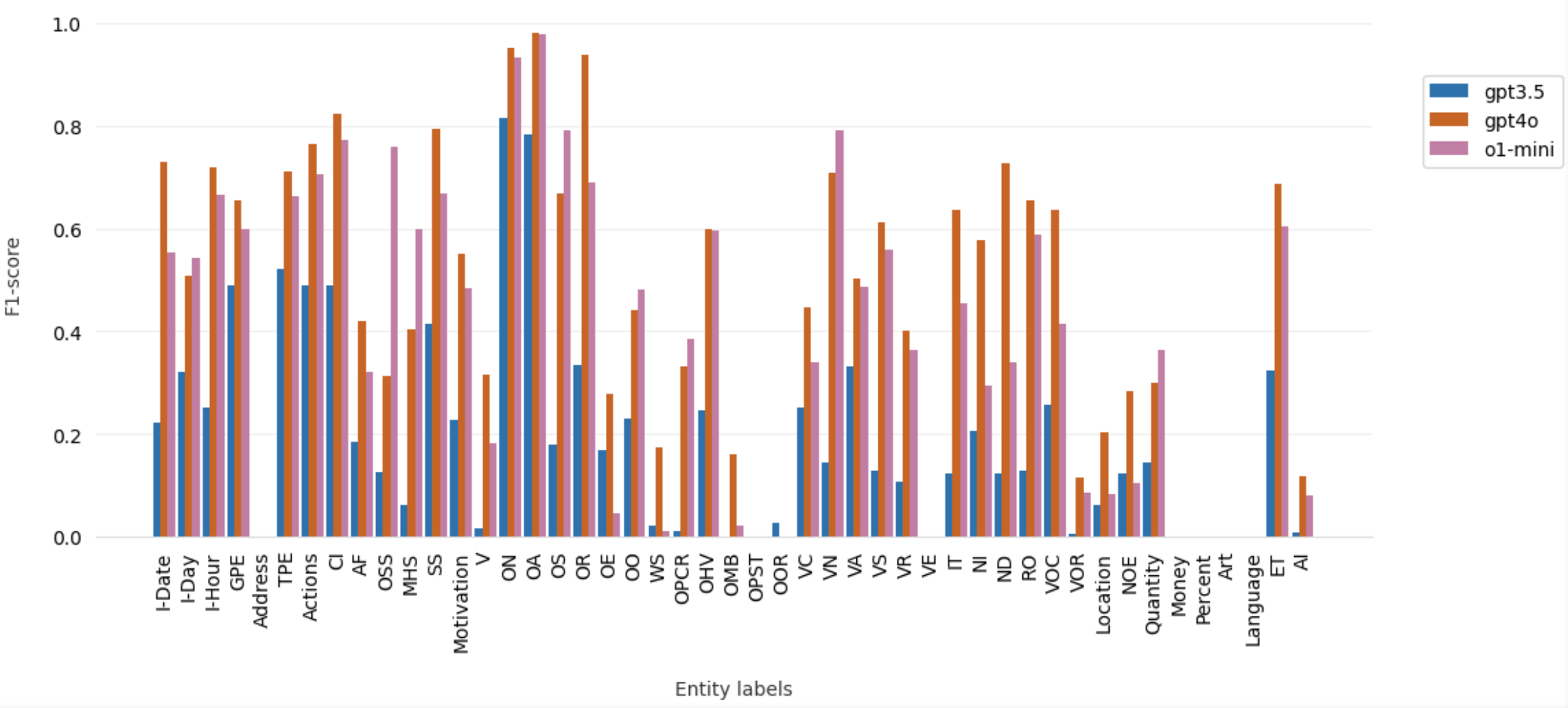}
    \caption{Performance comparison regarding F1-score across different types of named entities with 14 shots.}
    \label{fig:per}
\end{figure*}

Table~\ref{tab:results} presents the NER performance comparison on mass-shooting events using different shot counts of samples across GPT-3.5, GPT-4o, and o1-mini. It is observed that GPT-4o consistently outperforms both GPT-3.5 and o1-mini across all prompt sizes. It achieves the highest Micro Precision with various shots. For instance, with 10 shots, GPT-4o reaches 0.6903, significantly surpassing GPT-3.5 (0.6227) and o1-mini (0.6458). Similarly, GPT-4o maintains higher Micro Recall across all shots, reaching 0.6814 with 14 shots, which is notably higher than GPT-3.5 (0.3241) and o1-mini (0.6051). Furthermore, GPT-4o consistently achieves the highest Micro F1-score, demonstrating a better balance between Micro Precision and Micro Recall, with its best performance with 14 shots (0.6823).

Moreover, all models exhibit performance improvements as the number of shots increases, indicating that more context examples enhance NER accuracy. Both GPT-4o and o1-mini show significant gains in Micro F1-score as the number of shots grows, highlighting their better adaptability to few-shot learning. However, GPT-3.5 shows only marginal improvement in Micro Recall, rising slightly from 0.3062 (5 shots) to 0.3241 (14 shots). This suggests that GPT-3.5 struggles with recall, even when provided with more samples.

Despite being smaller and more resource-efficient, o1-mini performs competitively with GPT-4o in certain cases, particularly in Micro Precision. For example, with 14 shots, o1-mini achieves 0.6556 Micro Precision, which is close to GPT-4o's 0.6831. However, o1-mini's Micro recall remains consistently lower, indicating it may miss some relevant entities but still offers solid accuracy for detected entities.

In summary, GPT-4o performs as the most effective model for mass-shooting NER, achieving the highest Micro Precision, Micro Recall, and Micro F1-scores. Meanwhile, o1-mini demonstrates competitive performance, making it a resource-efficient alternative for less complex NER tasks. Increasing the shot count improves performance for all models, but the gains are more pronounced for GPT-4o and o1-mini, reflecting their superior adaptability to few-shot learning.

Figure~\ref{fig:per} illustrates the F1-score comparison across various named entities using 14-shot prompts. The results reveal that GPT-4o consistently outperforms GPT-3.5 and o1-mini in recognizing the majority of entity types. The orange bars (GPT-4o) achieve the highest F1-scores for most entity labels. Notably, for critical entities such as ON (Offender Name), OA (Offender Age) and OR (Offender Race), GPT-4o achieves over 95\% F1-scores, surpassing the other models. This demonstrates GPT-4o's stronger ability to recognize the critical entities with greater accuracy.

o1-mini delivers competitive performance for certain entity types. The pink bars (o1-mini) show F1-scores that occasionally approach or exceeds GPT-4o’s performance. For instance, for VN (Victim Name), OS (Offender Sex), MHS (Mental Health Issue), o1-mini outperforms GPT 4o's F1-scores performance after 14 shots, making it a resource-efficient alternative for these categories. However, its performance is less consistent across all entity labels compared to GPT-4o, indicating limited generalization to more categories.

In contrast, GPT-3.5 underperforms across most entity labels. The blue bars (GPT-3.5) show noticeably lower F1-scores across the board. For complex entity types such as OS (Offender Sex) and VC (Victim Count), GPT-3.5 achieves only around 0.2 - 0.5 F1-scores, significantly lower than GPT-4o and o1-mini. This reflects GPT-3.5's weaker contextual understanding and reduced adaptability to domain-specific NER tasks.

For minority or low-frequency labels such as AI, OMB, AF, GPT-4o and o1-mini exhibit a substantial advantage over GPT-3.5, achieving up to 4-times higher F1-scores. This indicates that larger models like GPT-4o handle rare or complex entities more effectively, likely due to their better generalization and few-shot learning capabilities.

\section{Related Work}
\label{sec5}

\subsection{AI-Driven Justice}
AI-driven justice leverages computational tools to enhance fairness, efficiency, and consistency in legal processes by automating evidence analysis, real-time crime detection, and predictive policing\cite{Leese03072024}. Previous studies indicate that predictive policing systems have reduced property crime rates by up to 13\% in U.S. pilot programs\cite{10452618}, while NLP models have achieved over 90\% accuracy in extracting critical case elements from court transcripts~\cite{brown2020language}. Recent advancements in LLMs such as GPT-4o and o1-mini enable efficient parsing of complex legal texts and accelerate tasks like legal document reviews by up to 40\%\cite{min2023artificial}. For instance, Freedman et al.\cite{wan-etal-2024-reformulating} demonstrate that domain adaptation of LLMs enhances legal NER accuracy, surpassing previous benchmarks on curated court filings. Similarly, Henderson~\cite{Galli_Sartor_2023} highlights recent improvements in predictive analytics, showing that newly released models exhibit enhanced interpretability, mitigating prior concerns regarding legal bias. Additionally, Liu and Kim~\cite{jayakumar2023large} present a systematic review of AI applications in legal dispute resolution, providing empirical evidence of cost savings and backlog reduction in pilot courts across various U.S. states. These AI systems uncover hidden patterns and relationships, making them essential for rapid insights in mass-shooting investigations and other justice-related applications.

\subsection{LLM-Based NER}
LLM-based NER has rapidly evolved, driven by the emergence of transformer-based models trained on large, diverse text corpora~\cite{devlin2018bert}. Previous studies demonstrate that GPT-4-inspired frameworks can increase recall by approximately 5\% over baseline models, enabling better generalization, particularly for rare entities. Recent benchmarks report F1-scores exceeding 93\%, indicating substantial gains in both precision and recall for practical applications{~\cite{wang2023gpt}. Contextual models, which excel at handling long documents, demonstrate a 10\% reduction in entity disambiguation errors, reflecting their improved capacity to parse complex texts. This enables them to reliably identify critical entities such as suspects, victims, and high-risk locations with over 90\% precision, as confirmed by FBI crime statistics~\cite{fbi2024}. The impressive performance of LLM-based NER stems from: 1) Enhanced AI-driven justice applications, enabling systems to rapidly adapt to domain-specific tasks; 2) Large-scale pre-training, which improves the model's ability to capture nuanced and context-rich details, making it highly effective for mass-shooting investigations and similar legal scenarios.

\section{Conclusion and Future Work}
\label{sec6}
This study implements knowledge acquisition on mass-shooting events using NER techniques to advance AI-driven justice. We leverage LLM-based NER with few-shot context learning to extract entities from real-world mass-shooting reports. Specifically, we employed several LLMs, including GPT-3.5, GPT-4o, and o1-mini, to identify entities related to mass-shooting incidents. Experimental results on real-world mass-shooting datasets demonstrate that GPT-4o outperforms other models in mass-shooting named entity recognition (NER), achieving the highest Micro Precision, Micro Recall, and Micro F1-scores. In comparison, o1-mini shows strong performance as well, offering a resource-efficient alternative for simpler NER tasks. The findings also suggest that increasing the shot count boosts the performance of all models, with GPT-4o and o1-mini benefiting the most, highlighting their exceptional adaptability to few-shot learning environments.

In the future, we aim to enhance NER performance on these minority categories by refining prompts. Additionally, we plan to develop advanced extraction techniques to capture relations between entities. It allows us to build comprehensive knowledge graphs of mass-shooting events, supporting more robust and actionable insights for justice-related applications.

%

\section*{Acknowledgment}
This research work is supported by NSF  under award numbers 2323419 and 2401860,  and by the Army Research Office under Cooperative Agreement Number W911NF-24-2-0133. The views and conclusions contained in this document are those of the authors and should not be interpreted as representing the official policies, either expressed or implied, of the NSF or the Army Research Office or the U.S. Government. The U.S. Government is authorized to reproduce and distribute reprints for Government purposes notwithstanding any copyright notation herein. Additionally, they acknowledge the use of AI-based tools, such as ChatGPT, for assistance in editing, grammar enhancement, and spelling checks during the preparation of this manuscript.



\bibliographystyle{IEEEtran}
\bibliography{References}
%



\end{document}